\documentclass[sigchi]{acmart}
\def\BibTeX{{\rm B\kern-.05em{\sc i\kern-.025em b}\kern-.08emT\kern-.1667em\lower.7ex\hbox{E}\kern-.125emX}}
    
\setcopyright{none}
\renewcommand\footnotetextcopyrightpermission[1]{} 
\setcopyright{acmlicensed}
\begin{document}

%
\title[Projection-based AV-to-Pedestrian Communication]{Designing for Projection-based Communication between Autonomous Vehicles and Pedestrians}

%
\author{Trung Thanh Nguyen}
\affiliation{
  \institution{School of Architecture, Design and Planning}
  \institution{The University of Sydney}
  \city{Sydney}
  \country{Australia}
}
\email{thanh.trung.nguyen@sydney.edu.au}

\author{Kai Holl{\"a}nder}
\affiliation{%
  \institution{LMU Munich}
  \city{Munich}
  \country{Germany}
}
\email{kai.hollaender@ifi.lmu.de}

\author{Marius Hoggenmueller}
\affiliation{
  \institution{School of Architecture, Design and Planning}
  \institution{The University of Sydney}
  \city{Sydney}
  \country{Australia}
}
\email{marius.hoggenmueller@sydney.edu.au}

\author{Callum Parker}
\affiliation{
  \institution{School of Architecture, Design and Planning}
  \institution{The University of Sydney}
  \city{Sydney}
  \country{Australia}
}
\email{callum.parker@sydney.edu.au}

\author{Martin Tomitsch}
\affiliation{
  \institution{School of Architecture, Design and Planning, The University of Sydney}
  \institution{CAFA Beijing Visual Art Innovation Institute, China}
}
\email{martin.tomitsch@sydney.edu.au}
%
\renewcommand{\shortauthors}{Nguyen et al.}

%
\begin{abstract}
Recent studies have investigated new approaches for communicating an autonomous vehicle's (AV) intent and awareness to pedestrians. This paper adds to this body of work by presenting the design and evaluation of in-situ projections on the road. Our design combines common traffic light patterns with aesthetic visual elements. We describe the iterative design process and the prototyping methods used in each stage. The final design concept was represented as a virtual reality simulation and evaluated with 18 participants in four different street crossing scenarios, which included three scenarios that simulated various degrees of system errors. We found that different design elements were able to support participants' confidence in their decision even when the AV failed to correctly detect their presence. We also identified elements in our design that needed to be more clearly communicated. Based on these findings, the paper presents a series of design recommendations for projection-based communication between AVs and pedestrians. 
\end{abstract}

\copyrightyear{2019}
\acmYear{2019}
\acmConference[AutomotiveUI '19]{11th International Conference on Automotive User Interfaces and Interactive Vehicular Applications}{September 21--25, 2019}{Utrecht, Netherlands}
\acmBooktitle{11th International Conference on Automotive User Interfaces and Interactive Vehicular Applications (AutomotiveUI '19), September 21--25, 2019, Utrecht, Netherlands}
\acmPrice{15.00}
\acmDOI{10.1145/3342197.3344543}
\acmISBN{978-1-4503-6884-1/19/09}

%
%


%

%

%
\maketitle

\section{Introduction}

Urban robots are currently making the transition from laboratory and test-bed environments to being evaluated in real-world urban contexts. This transition is driven by public authorities and industry, mainly with the aim to automate public services, logistics and transportation. A main focus is the development of the driverless car, leading to speculation and debate about how this technology may shape our cities in the near future \cite{Gandia2019}. One consideration which has seen much attention by the community recently, is how driverless cars and other autonomous vehicles\footnote{Some scholars have argued for using the term ``automated vehicles'' as such vehicles are not (yet) strictly autonomous. We use ``autonomous'' in this paper as it is, at the time of writing, still widely used in previous research studies and in the industry to refer to these types of vehicles, and since our research focuses on future scenarios where vehicles might indeed be fully autonomous.} (AVs) communicate their internal state to nearby pedestrians \cite{Mahadevan:2018:CAI:3173574.3174003, Rasouli2018AutonomousPractice}. This is an important challenge as there is no human driver that is able to interact with pedestrians \cite{article} and for AVs operating safely in an urban environment, effective communication channels are critical to help pedestrians in making decisions, e.g. about when and where to cross a road. To address this challenge, various design concepts have been developed by academic research, design studios and the automotive industry. For example, the London-based design studio Umbrellium developed an interactive LED-based pedestrian crossing\footnote{\href{http://umbrellium.co.uk/initiatives/starling-crossing/}{http://umbrellium.co.uk/initiatives/starling-crossing/}; \textit{last accessed: April 2019}} that responds dynamically to cars and pedestrians. However, as scaling up such design concepts would require the upgrade of existing city infrastructure, resulting in high costs, others focused on using the external surface of cars as displays \cite{Colley2017}, ranging from embedded low-resolution LED displays \cite{Gandia2019}, on-car projections \cite{Colley:2018:CES:3205873.3205880} to attached screens\footnote{\href{https://www.drive.ai/}{https://www.drive.ai/}; \textit{last accessed: April 2019}}. Using the car itself as a carrier to display relevant information for surrounding pedestrians, \citet{Colley:2018:CES:3205873.3205880} made the case for designing cars as a shared resource. However, as cars have a major influence on the cityscape, it is important that such displays on cars are not only designed for maximised efficiency and attention, but are also aesthetically pleasing and not contributing towards the increasing oversaturation of public spaces with pervasive displays \cite{Davies2018}.

To address this challenge, we present the iterative design of in-situ projections to help pedestrians in making crossing decisions. Projections were chosen as a suitable display technology as: (1) they are visible to a large number of pedestrians; (2) they can be pointed towards the street, which is already used as a surface for traffic signage, such as pedestrian crossings, thus providing a display location that pedestrians are familiar with; and (3) they can be easily switched off when not needed. While previous work on AV-to-pedestrian displays mainly focused on the effective encoding of information, in this paper we present methods for designing aesthetic aspects, treading the fine line between efficiency, safety and aesthetics. Throughout the design process, we developed prototypes of various fidelity levels, including sketches, animated mock-ups and interactive virtual reality (VR) simulations. We tested the final design concept in a VR user study, thereby focusing on a street-crossing scenario with driverless cars, to evaluate how AVs can communicate their intent and awareness to pedestrians via projections, especially in situations that are not facilitated by traffic lights. Testing the final design in VR enabled us to also evaluate the influence of sensor failures within the prototype on participants' confidence regarding the visualisations, in a safe, risk-free environment.

The contribution of this paper is three-fold: (1) We report on our approach of designing projection-based information cues using prototyping methods of various fidelity levels; (2) we present insights and findings from a VR simulation study evaluating the efficiency of those cues in terms of their comprehensibility with potential users; and (3) based on those findings, we derive design recommendations for projection-based communication between AVs and pedestrians.




\section{Related Work}
The following areas of research form the foundation for the study presented in this paper: (1) research on pedestrian behaviour when crossing a road, (2) publications addressing the communication between AVs and pedestrians, and (3) work that used projections as a communication channel.

\subsection{Pedestrian Crossing Behaviour}
Previous research stressed that the main source of motivation for pedestrians to cross a road in front of a vehicle is the size of the gap between approaching cars and the car's behaviour (e.g. accelerating or breaking)~\cite{yannis2013pedestrian, 8241847, Dey:2017:PIV:3122986.3123009}. Amongst others, Beggiato et al.~\cite{Beggiato:2017:GAT:3122986.3122995} investigated what size a gap should have and at what moment an autonomous vehicle should break in order to create a natural perception of driving behaviour from a pedestrian's perspective. A longer observation of oncoming traffic indicates pedestrian's willingness to cross a road~\cite{Rasouli_2017_ICCV, 7995730}.
After a decision is made, communication between a driver and pedestrians is an additional crucial aspect to foster safety and comfort~\cite{7995730, 8569324}. For example, pedestrians can provoke human drivers to come to a halt through staring~\cite{article} or gesturing~\cite{ZHUANG2014235}.

\subsection{AVs and Pedestrian Communication}
As a consequence of increasing automation in vehicle control systems, drivers are able to engage in non-driving-related activities~\cite{Pfleging:2016:IUN:3012709.3012735, doi:10.1080/15472450.2017.1291351}. Hence, advanced driving-assistance systems with an SAE level higher than three do not require ``drivers'' to interact with their surroundings~\cite{J3016_201806}. However, pedestrians benefit from information about the awareness and intent of highly automated vehicles, especially during crossing decisions~\cite{Mahadevan:2018:CAI:3173574.3174003, 7745210, 10.3389/fpsyg.2018.01336}. To overcome the lack of driver-to-pedestrian communication, research and industry developed various concepts to help with this~\cite{8667866}. As of now, most prototypes feature one or more external displays, with the display technology being embedded or attached to the outer shell of the vehicle~\cite{clamann_michael_aubert_miles_cummings_l_2016, DBLP:journals/corr/FridmanMXYFR17, inproceedings}. However, using conventional screens comes with the limitation that the content is only visible from a specific angle, which limits the number of pedestrians that can receive the information. The issue of multi-user support also accounts for externally attached contraptions, for example, additional eyes which follow pedestrians~\cite{Chang:2018:VSC:3239092.3265950, Chang:2017:ECI:3122986.3122989, Mahadevan:2018:CAI:3173574.3174003}. To enable communication with multiple users, approaches featuring personal devices such as smartphones have been tested as an alternative to displays on the car~\cite{Mahadevan:2018:CAI:3173574.3174003, nissan}. Finally, researchers stressed that AV-to-pedestrian communication channels need to be designed following a user-centred design process~\cite{mci/Holländer2018}, which is challenging as prospective users cannot yet draw on experiences interacting with AVs \cite{Frison:2017:WUD:3131726.3131734}.

\subsection{Projection-Based Concepts}
A promising approach is the use of projections on the road as this provides a display surface that is only restricted by the distance of the projector to the ground and the utilised hardware, rather than the design of the car itself, which limits the display's size and the deployment location. Further, projections are capable of presenting information in a wide range of visual encodings, from more ambient representations, such as colour and patterns, to symbols and high-resolution images. Mercedes Benz presented a prototype\footnote{\href{https://www.mercedes-benz.com/en/mercedes-benz/innovation/research-vehicle-f-015-luxury-in-motion/}{https://www.mercedes-benz.com/en/mercedes-benz/innovation/research-vehicle-f-015-luxury-in-motion/}; \textit{last accessed: March 2019}} which is able to project a crosswalk on the ground to yield the way for non-motorised road users. Mitsubishi\footnote{\href{http://www.mitsubishielectric.com/news/2015/pdf/1023.pdf/}{http://www.mitsubishielectric.com/news/2015/pdf/1023.pdf/}; \textit{last accessed: March 2019}} uses projection to indicate a vehicle's intended direction (straight, back, left or right). Colley et al.~\cite{Colley:2018:CES:3205873.3205880} explored projections on the car and ground to remind drivers about forgotten items (e.g. phone and keys) or to display safety information and intentions outside of the vehicle. In their study, displaying safety-critical information to pedestrians was perceived as most beneficial and received social acceptance. Dancu et al.~\cite{dancu} applied a projection of directional intentions to a bicycle and showed that their system simplified predicting movements for other road users. A common shortcoming of projection-based systems is their limited visibility during bright daylight. This limitation is increasingly addressed through the availability of laser projection systems. To account for visibility issues, we designed our system and study in a dusk scenario. In future implementations, projection-based solutions could also take on a complementary function to support other communication systems in poor lighting conditions. Uneven road surfaces (e.g. gravel) or interference of multiple vehicle projections might hinder the perception of signals. However, these challenges go beyond the scope of the aims of our study and offer avenues for future research.

\section{Design Process}
In this section, we report on the iterative process of designing in-situ projections on the road for safety critical situations. As the context - designing for communication channels between autonomous vehicles and pedestrians - is still rather underrepresented in the field of human-computer interaction (HCI), we could not fall back on existing purpose-built methods and tools. We therefore applied common interaction design approaches and tailored them to our specific context.

\subsection{Situation}

We designed the projection-based visualisations for the situation in which a pedestrian intends to cross the street at a pedestrian crossing and sees an AV approaching. In this situation, supporting information cues are of importance as the traffic flow is not facilitated by a traffic light and interactions between the driver and pedestrian are not possible. This means that the pedestrian has to solely rely on the AV's sensors, which can lead to uncertainty in the pedestrian's decision to cross the road. Furthermore, it has been reported that 25\% of all accidents in Europe that resulted in the death of pedestrians happened at a pedestrian crossing \cite{article}.

Therefore, the crossing scenario has been well researched (e.g. \citet{7745210}), however, not with a projection-based design. Applying our projection concepts to the crossing scenario allows us to discuss our results in relation to the insights from previous research that used other display solutions. Previous research suggested that malfunction and errors reduce people's reliance on automation systems \cite{Yang2017EvaluatingAutomation}. In this study, we aimed to understand to what extent pedestrians still rely on information cues from AVs - to make decisions and ensure their safety - even when the vehicle's sensors are malfunctioning (e.g. not detecting the participant).


\subsection{Projection Concepts}

Designing the projection concepts for the crossing situation, we followed an iterative design process, including low-fidelity sketches, digital mock-ups, animated 2D prototypes in the java-based programming environment Processing \footnote{\href{https://www.processing.org}{https://www.processing.org}; \textit{last accessed: April 2019}} and the final 3D prototype developed in Unity\footnote{\href{https://www.unity.com}{https://www.unity.com}; \textit{last accessed: April 2019}}. We used low-fidelity prototyping methods to quickly explore a wide range of ideas and concepts. 2D prototypes were then developed to refine the design concepts and to derive specifications for the visual appearance of the projections (e.g. visual elements, animation sequences). 

\begin{figure*}[t]
 \centering
  \includegraphics[width=2.1\columnwidth]{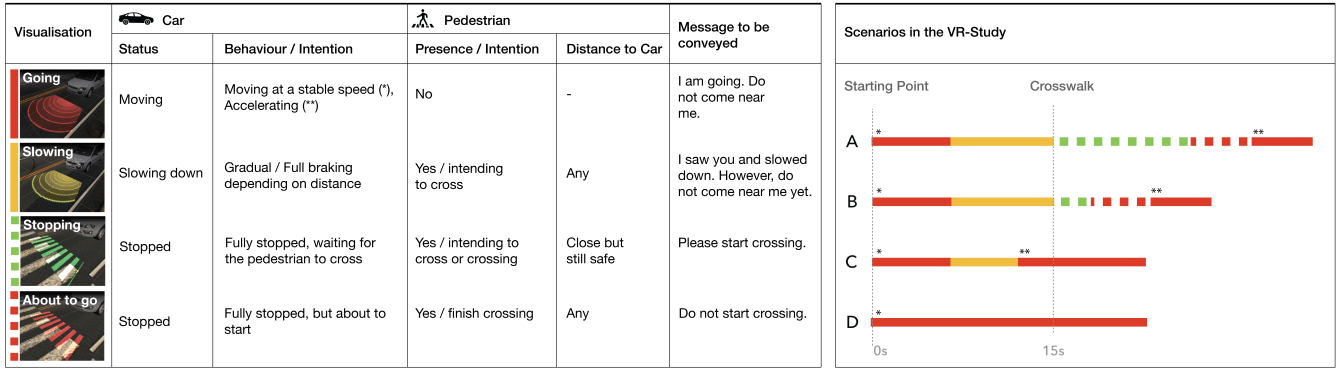}
  \caption{The four key visualisations signifying the intent and awareness of the AV in an ideal scenario (left), and the four scenarios used in the VR study, including the ideal scenario (A) and three scenarios with varied degrees of sensor failures (B, C, D) (right).}
  \label{fig:study_design}
\end{figure*}

We adapted the template for mapping interaction between driverless cars and pedestrians from the framework created by Owensby et al.~\cite{Owensby2019}. In our chosen situation, a vehicle's ideal response to prioritise pedestrian safety is to slow down, stop in front of the crossing lines and wait until all pedestrians finish crossing before starting again. We broke down this sequence to the vehicle's status, its awareness of pedestrians, pedestrian's intent to cross and the distance between the vehicle and pedestrians. Using these parameters, we define four stages of the AV in the crossing situation as \textit{moving, slowing down, stopping,} and \textit{about to go} (See table on the left in Figure \ref{fig:study_design}). Owensby et al.'s framework highlights the need for pedestrians to know the current status of the car, its intent and whether it is aware of them. We added an additional consideration (``Message to be conveyed'') to interpret the meaning of our visualisation from an observer's point of view. We then designed four types of visualisations corresponding to the four statuses of the AV. A storyboard was developed with preliminary visualisations to understand how they can support the interaction flow in the contextual situation. Figure \ref{fig:study_design} shows the revised, final high-fidelity visualisations for each status of the AV.

\subsubsection{Ideation}

\begin{figure}[!hb]
  \includegraphics[width=\columnwidth]{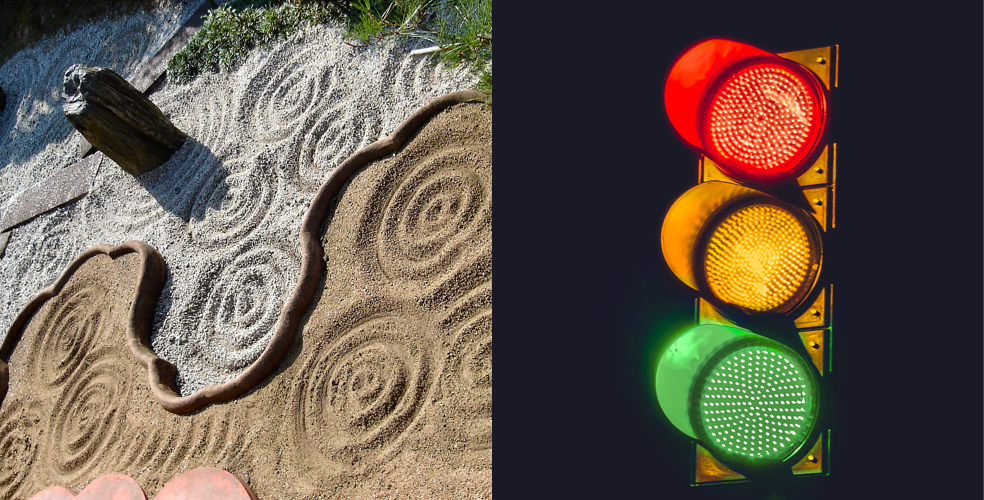}
  \caption{(Left) Japanese Zen garden with gravel that mimics wave pattern, creating a calm feeling. Image credit: PlusMinus via Wikimedia. (Right) A traffic colour scheme was used in our design solution to take advantage of existing traffic communication practices. Image credit: Harshal Desai on Unsplash.}
  \label{fig:inspirations}
\end{figure}

We sketched various ideas for projection visualisations on paper. This included text as Chang et al.~\cite{Chang:2018:VSC:3239092.3265950} found that presenting text information was preferred by participants when communicating a vehicle's intention compared to other modalities (eyes on a car, smile icon, lights and projections). However, using text has limitations over symbols in particular in multi-cultural environments where multiple languages are spoken. Furthermore, text excludes illiterate people and children that are not yet able to read, reading text requires the foreground of user's attention, and readability is hampered when the car is in motion. Another study also found that people do not interpret text quickly enough in this context \cite{clamann_michael_aubert_miles_cummings_l_2016}. These considerations led us to using aesthetic ambient lighting displays as a reference, as these types of displays are typically designed to communicate information to people in a low cognitive load approach. In aesthetic ambient lighting displays, the use of natural phenomena, such as ocean waves to visualise data of a system, has been proven to attract attention subtly and help observers understand how the systems works \cite{Jones2018YourCamera,Shelton2017EvaluatingTraffic}. Sonar, a method to communicate or detect objects in the water \cite{Ainslie2010}, was the main inspiration for the wave pattern in our projection-based design concept. As part of human verbal communication, sound when visualised can highlight the expressive content of the communication channel \cite{Pietrowicz2013Sonic:}. The wave pattern also appears in Japanese gardens, which exhibit a calming atmosphere \cite{VanTonder2005}. The gravel with wave shapes (see Figure \ref{fig:inspirations}) around the objects highlights the presence of objects \cite{VanTonder2005}.

In the following step, we converted the sketches to digital mock-ups to investigate variations of colours, shapes and the visualisation's relative size to the vehicle through many rounds of iteration. One of the challenges we had to overcome with our design proposal was the limited number of visual elements to make four types of visualisations distinguishable in reflection of the four states of the AV. We used three visual elements in our visualisation: colours, patterns and movement created by shifting patterns and colour. The traffic light colour scheme was mapped to the car's status. We used a wave pattern for when the car was still moving and crossing lines for when the car had stopped and it was safe for the pedestrian to cross in front of the car. The combination of colours, patterns and movements allowed us to create distinctive visualisations to communicate the car's intent and awareness of pedestrians in all four states. 

\subsubsection{Dynamic Prototype}

Processing was used as a medium to prototype the patterns and to fine-tune the visual aspects such as size, shape and movement. The speed of radiating waves and chasing green crossing lines in this prototoype was controlled manually through keyboard input and mapped to the vehicle's speed. 

\subsubsection{Virtual Reality Prototype}

We created the final prototype representation in VR, allowing us to adopt VR as a testing method, similar to previous studies of AV-to-pedestrian communication systems (e.g. \cite{Pillai2017, Mahadevan:2018:CAI:3173574.3174003}). While previous studies asked people to indicate whether they would start crossing the street by pressing a button (e.g. \cite{Mahadevan:2018:CAI:3173574.3174003}), we aimed to implement a situation that was closer to a real-world situation. To that end, we focused on achieving visual realism, by using a high frame-rate, field of view, head tracking system and position tracking in VR, which has been demonstrated to enhance realistic behavioural responses \cite{Slater2009}. Testing our solution in the VR environment also allowed us to put participants in risky situations which would not be safe to evaluate in real-world settings.

The prototype was created using Unity and 3D models. The vehicle is a 5-seater midsize sedan model. The wheels had animation that closely matched the car's speed and turning movement. We did not model any person inside the vehicle, and improved the lighting condition so that participants could observe the absence of a human driver.  The environment was created using pre-made models of an urban environment with a resemblance to a suburb familiar to our participants. This was chosen to create a comfortable feeling by providing a sense of familiarity in the scene. The daylight was adjusted to create long shadows in the initial view of participants, supporting visual realism. The overall brightness was dimmed to match the dusk time and to be more suitable for visibility of the projection on the ground. A HTC Vive VR headset was used to experience this environment.

We developed a logic to control the car speed, braking and turning of the wheels according to the driving path and the visibility of a human's 3D object in the environment. The status of the car was communicated via the projection in such a way that a change of the cars' status triggered a change of the displayed information. We developed a test control of the car so that we could disable the car's sensor anytime, which allowed us to create deliberate malfunctions in some of our study scenarios. 

\section{Evaluation Study}

\subsection{Experiment Design}
As our research objective was to study how people perceived the vehicle's intent and awareness via projection-based communication in different situations (including malfunctioning scenarios where the sensor would fail to ``see'' the pedestrian, e.g. due to poor light conditions), we prepared four scenarios (See Figure \ref{fig:study_design}, diagram on the right). This allowed us to evaluate how pedestrians would react and behave across different scenarios and whether the communication channel enables people to make safe decisions. The scenarios are:
\begin{itemize}
\item Scenario A: Car sensor works correctly throughout the scenario.
\item Scenario B: Car sensor works correctly at first and detects the pedestrian on the side of the road, but it fails to correctly detect the pedestrian while they are crossing.
\item Scenario C: Car sensor works correctly initially, detecting the pedestrian and making the car start to slow down, but then fails to detect the pedestrian correctly before the car completely stops.
\item Scenario D: Car does not detect the presence of the pedestrian at all.
\end{itemize}

Factors that influence pedestrian behaviour in a crossing situation are driving speed and gap distance between the pedestrian and the car \cite{clamann_michael_aubert_miles_cummings_l_2016}. We kept these control variables constant by having the same street and the same AV coming to the crosswalk from the same side in all scenarios.

\subsection{Procedure}

As a sense of presence in VR is essential for participants to display realistic responses \cite{Slater2009a}, we followed the suggestions made in previous research \cite{Slater2002} to have different activities before the test, such as asking people to walk around while wearing the VR headset. Body movement, such as walking, can be tracked in a virtual environment and can help produce a sense of presence more than pointing and clicking \cite{Slater2002}.

In our experiment, the participant was placed on the footpath near a marked pedestrian crossing. Before starting the test scenarios, the participant was asked to practice crossing the street when there were no cars approaching. This was to help participants become familiar with the environment. We also asked several questions about the scene to help participants adapt their visual perception to a new environment. Next, we started a familiarisation scene, in which the participant stood on the pavement at some distance from the road to observe cars passing by without having to make a crossing decision.  The cars throughout this scene functioned properly and only displayed the \emph{Moving} visualisation as the pedestrian was too far from the road to be detected as intending to cross the road. As part of this scene, we also asked participants if they could identify the cars as driverless, prompting them to look closely in order to realise that there was no driver sitting behind the steering wheel. This was important to ensure participants were aware that their decision to safely cross the road could not rely on interpersonal communication. 

Before starting the first scenario, the participant was asked to adjust their standing position on the pavement until they felt comfortable and safe. We then started the first test scenario. The sequence of the four scenarios was randomised to counterbalance any potential learning effects. After each scenario, the participant took off the headset and sat down in a nearby booth to complete a short interview and questionnaire. After all four scenarios were completed, the participant was asked to complete a post-study interview about the overall experience in VR and to compare all four scenarios. 

\subsection{Measurement}



\begin{figure}[h]
  \includegraphics[width=\columnwidth]{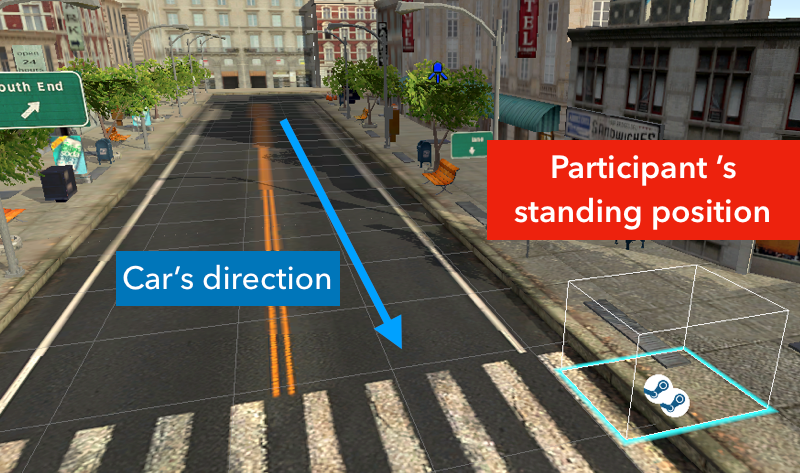}
  \caption{The scenario set-up in VR: The car starts its trajectory at the corner of the street and approaches the pedestrian from the right side.}
  \label{fig:vr_setup}
\end{figure}


We used the following methods to collect both qualitative and quantitative data about participants' behaviour, perception and feedback:

(1) Video recordings: Participants' behaviour (e.g. stepping forward or stepping backward) while wearing the VR headset were recorded via a video camera. The video recordings were later used to verify participants' statements during the data analysis. Further, people have their own crossing strategies \cite{clamann_michael_aubert_miles_cummings_l_2016}, making it difficult to compare participants' response times across scenarios. 

(2) Post-scenario semi-structured interview questions: After each scenario, participants were asked to answer a semi-structured interview to capture their understanding of the visualisation. The interview questions were adapted from previous studies \cite{Pillai2017,Owensby2019} and included questions about participants' understanding of the vehicle's behaviour and which factors they used to make the crossing decision. Each interview was between 5 and 10 minutes and audio-recorded for later transcription.

(3) Post-study semi-structured interview questions: After completing all four scenarios, participants were asked about their overall experience and to compare the four scenarios. The aim of this interview was to understand how participants understood and learnt the vehicle's behaviour. Each interview lasted between 10 and 15 minutes and was audio-recorded for later transcription.

We did not record the time it took pedestrians to respond as we did not compare different visualisation approaches, which has been the focus in previous AV-to-pedestrian interface studies (e.g.  \cite{clamann_michael_aubert_miles_cummings_l_2016}). Instead, our aim was to understand the participants' reactions in each of the four different scenarios (e.g. car slowing down versus not slowing down). 


\subsection{Participants and Location}
Our study involved 18 participants (between 18 and 44 years; 11 female, 7 male; students and working professionals). Out of our participants, 15 had previous experience with VR (from social to gaming applications), possibly because our call for participants stated that the study involved a VR experience. Three participants were working in the research fields of smart cities and autonomous vehicles; one of them had interacted with real AVs through their research.

In terms of visual acuity, 13 participants had normal vision, the other 5 had to wear glasses to correct their eyesight. Participants with short-sighted vision were asked to wear their usual glasses throughout the study. All participants were able to read, listen and speak fluently in English. None of the participants had any mobility impairment. All participants read the Participant Information Statement (approved by our university's ethics board) and signed the consent form to permit us to record video, photos and audio. 

The study was carried out in our lab, which offers a free space of about 6 by 6 metres. The tracked space through the VR sensors was not large enough for participants to completely cross the entire street in VR, but it was sufficient for participants to step from the footpath onto the road to signal their intention to cross.

\begin{figure}[t]
  \includegraphics[width=\columnwidth]{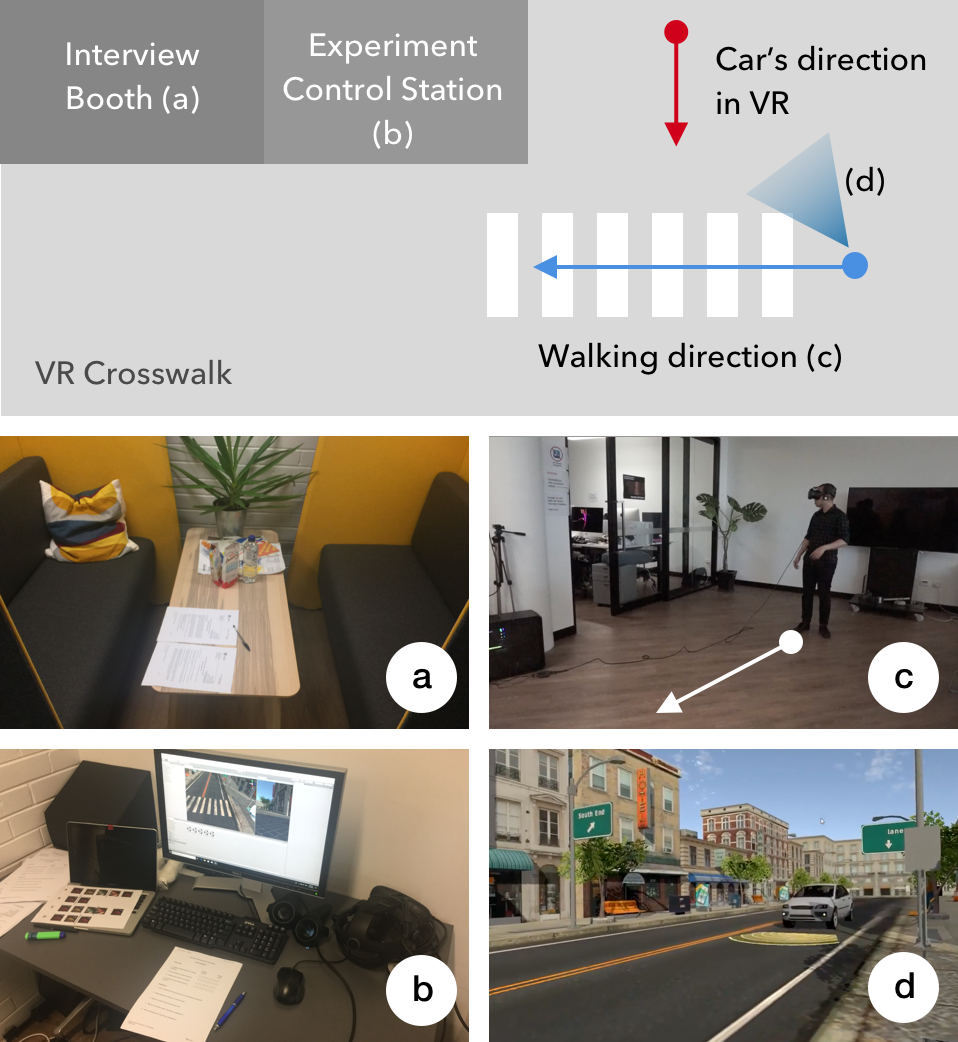}
  \caption{The study environment and setup, including the (a) interview booth; (b) control computer; (c) VR space; and (d) virtual space.}
  \label{fig:study_location}
\end{figure}

\subsection{Pilot Study}

Prior to our main investigation, we conducted two pilot studies (3 participants each, who did not participate in the main study) to refine the study setup and materials. The first pilot study involved participants wearing the headset throughout all four scenarios including each of the post-scenario interviews. Our intention was to allow participants to remain immersed in the virtual environment while completing the interviews, however, we found that this increased the levels of discomfort and risk of nausea because of the study duration. In the second pilot study, to increase the comfort of participants, we asked them to take off the VR headset and complete the post-scenario interview while sitting in the interview booth. The second pilot study also allowed us to test and revise the way we recorded the video and how the instructions were delivered to participants. Afterwards, we conducted the main study with 18 participants by following the aforementioned procedure.   

\section{Results and Discussion}

For the analysis, we transcribed the interviews that we conducted with the participants after each scenario. We then analysed the transcriptions using affinity diagramming in order to sort the data into similar concepts (clustering) and identify themes \cite{Lazar:2010:RMH:1841406}. 
Upon transcribing statements made in the interviews that involved an action by the participant (e.g. stepping back), we used the video recordings to verify these statements. If the interview statement did not match the observation in the video, we excluded the respective data. 
In the following, we discuss the findings of our qualitative data analysis, including a set of design recommendations (DR) (see Table \ref{tbl:design-recommendations}).

\begin{table}[b]
  \includegraphics[width=\columnwidth]{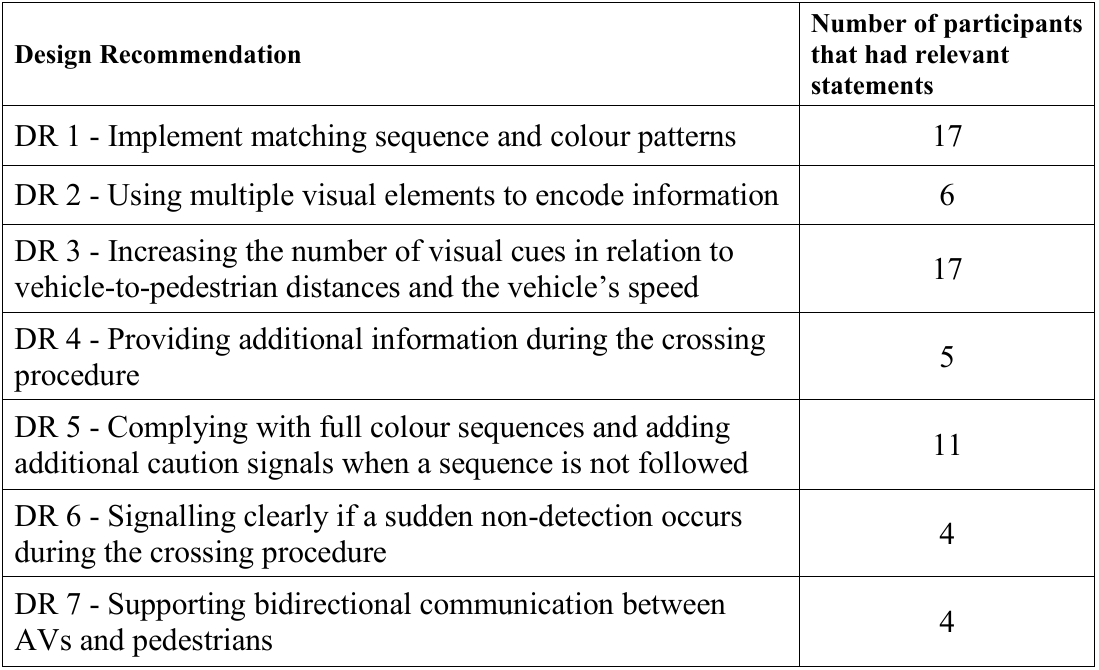}\Description{excel-table} 
  \caption{List of design recommendations.}
  \label{tbl:design-recommendations}
\end{table}

\subsection{Interpretation of Colours}
In our design, colours were mapped to the vehicle's status and intent. The interviews confirmed our observations in the sense that most participants acknowledged this association after performing all four scenarios: in total, 17 participants stated that in their view a change of colour signalled a change of the vehicle's behaviour. However, as previously reported \cite{10.1007/978-3-319-60441-1_63} pedestrians might not necessarily understand whether the displayed information is an instruction or expresses the vehicle's intentions. Commenting on the hesitation to cross when seeing the green crossing visualisation in the first scenario, participant P8 said \textit{``(At first) I think it is indicating that I shouldn't go when it is red. But when the light is in front of the car, it felt that the light is [an] indicator for the car, not for me''}. In the post-study interview, P6 said \textit{``I don't know who is the object of the light. The traffic light is usually for the driver.''}
Participant P8 commented on the straight crossing bars changing from green to red colour: \textit{``The light changed. The radar changed from green to red so [the vehicle] indicated that it was about to start the engine.''} 
One participant (P2) misunderstood the projections in the sense that she referred to them as the \textit{``normal lighting of the car''} and concluded that therefore \textit{``there is no significance''}. She mentioned this in the post-study interview, even after finishing all scenarios, including the one in which the car's sensors worked correctly. As a consequence, she used the car's speed as the main factor to make a decision at the crosswalk. For participants that interpreted the colour encoding correctly, the speed of the vehicle was an important factor to reinforce the interpretation of meaning: \textit{``[The vehicle] slowed down and [...] the process of changing colours makes me feel that it changed [colours] because of me being there [...]''} (P9). Another participant stated: \textit{``Yellow means [the vehicle] is slowing down, noticing something and checking for caution''} (P8). The real-time synchronisation between the car's speed and the change of colour was perceived by the participants in the sense that the vehicle was trying to detect or already detected their presence. This was articulated by all participants after they finished the study, though the level of certainty varied. People were most certain about the car's awareness of them when the car stopped.\textit{``The car stopped so it must have noticed I am here''} (P8).\\

\subsubsection{Design Recommendation (DR) 1}
\textit{\textbf{Implement matching sequence and colour patterns:}}
In line with previous research, we found that the car's kinematics are an important factor for pedestrians to make a crossing decision~\cite{Dey:2017:PIV:3122986.3123009}. Therefore, additional visual design elements should match their sequence with the car's motion in order to reinforce people's existing understanding and foster intuitiveness. Furthermore, adopting the well-known colour indications from traffic lights has shown to help people adapt quickly to the sequential stages of the car's behaviour. However, a limitation seems to be that pedestrians need to experience the full colour spectrum at least once. Two of the participants who experienced the successful scenario at the end of the experiment mentioned that the red colour would be misleading and concluded that only after seeing the full sequence of red-yellow-green they did understand the meaning of the red colour. Referring to scenario D, where the car did not stop at all, one participant stated that the red colour \textit{``doesn't look safe''}. However, this participant experienced the situation rather in the sense that \textit{``the car is just passing by and it's not that dangerous''} (P5). Another participant (P6) mentioned instead: \textit{``I will be cautious with the red colour, especially when I don't get to know about the yellow and green [colour]''}.\\


\subsection{Interpretation of Crossing Patterns}
While the majority of the colour encoding was the primary source to make sense of the visualisations, two participants also referred to the crossing patterns in more detail. When being asked which factors informed them about the vehicle's intention, they described the shape and movement of patterns closely to the design: \textit{``The pattern changed as well, from concentric circles to parallel bars'' (P12)}. Another added that the crossing patterns were a helpful way to provide additional directional information at the crossing: \textit{``The pattern changed from circles to crossing lines. It moved and told me to go in a certain direction. It felt like [the visualisation] was leading me to go across''} (P9). In general, those subtle movements of light were rather perceived as an aesthetic ``add-on'', with one participant (P15) mentioning \textit{``how the wave is transforming, from lighter to darker colour [...], is visually very attractive''}.\\

\subsubsection{DR 2}
\textit{\textbf{Using multiple visual elements to encode information:}} Though encoding safety-critical information using colours seems to be most appropriate for the majority of users, for others the pattern design also played an important role in their decision-making. Under different conditions, people might not see the change of colours correctly so the differences in movement, shapes and size should be used as a redundancy measure. This consideration will be also helpful for people with red-green visual impairment, guaranteeing inclusiveness in the design.\\

\subsubsection{DR 3}
\textit{\textbf{Increasing the number of visual cues in relation to vehicle-to-pedestrian distances and the vehicle's speed:}} From far distances, people are most sensitive to changes in colour. At closer distances and when the vehicle slows down, more detailed visual elements, such as patterns and animations, or even explicit information, such as text, can be perceived.\\

\subsubsection{DR 4}
\textit{\textbf{Providing additional information during the crossing procedure:}} Pedestrians feel more confident if they receive information about the time they have to finish the crossing. Additional information, in form of a timer, could be visualised using an explicit layer of text (re DR 3). Though under optimal conditions, the autonomous vehicle would wait until pedestrians have finished the crossing procedure, additional real-time visualisations indicating that the vehicle is still aware of the pedestrian's presence, could increase the perceived trust.\\

\subsection{Catering for Pedestrian Non-Detection}
Participants expressed the need to receive more detailed information about the vehicle's decision to start the engine again. Even when they followed the instructions correctly and made safe decisions to cross the road, they were uncertain what triggered the car to change the visualisation. After experiencing scenario B when the car sensor failed in the middle of the crossing procedure, two participants (P5, P7) expressed their concern about the uncertainty of how long they had time to finish the crossing, stating: \textit{``What makes me feel unsafe is how long the car would be waiting for me''} (P5); and \textit{``I'm not sure if I have enough time to go back to the safe area''} (P7).\\

\subsubsection{DR 5}
\textit{\textbf{Complying with full colour sequences and adding additional caution signals when a sequence is not followed:}} In malfunction scenarios, when the full visualisation sequence (i.e. red-yellow-green) is not followed correctly, pedestrians can be confused due to not having seen the full sequence of signals (re: DR 1 - traffic light scheme). Therefore, we suggest that additional visualisations need to be designed for scenarios in which the full sequence is not followed (e.g. due to sensor malfunctioning). For example, if the visualisation sequence transitions from slowing down (yellow) to accelerating (red), an additional visual cue should be added to inform pedestrians about potential system failures.\\

\subsubsection{DR 6}
\textit{\textbf{Signalling clearly if a sudden non-detection occurs during the crossing procedure:}} Building on DR 5, in case of a sensor failing to detect the pedestrian after successfully having stopped for a pedestrian, the car should not immediately start accelerating. To ensure the safety of pedestrians, it should display a countdown timer - possibly supported through audio cues - to signal its intention to any pedestrian that might potentially still be attempting to cross.

\subsection{Pedestrian's Negotiation with Vehicles}

Previous research has demonstrated the current methods of communication with autonomous vehicles such as using hand gestures or stepping forward to communicate the intention to the cars \cite{Sucha2017}. Such behaviour was observed in this study as well. \textit{``I stepped down to signify that I wanted to cross''} (P10). Participants also felt curious about how the car identified whether they were going to cross. \textit{``Why didn't it change back to green when I made clear that I want to cross this time.''} (P5 tried to step forward to signal her intent of crossing to the car)

Participants did not appreciate the AV giving them commands. For instance, \textit{``I feel like the car was commanding me to do things. I don't like that.''} (P14). Pedestrians also mentioned that there was \textit{``no negotiation''} (P2, P6) between them and the AV. This was due to having no real-time feedback to respond to a participant's gesture when they intended to cross.\\

\subsubsection{DR 7}
\textit{\textbf{Supporting bidirectional communication between AVs and pedestrians:}} The AVs need to be able to sense and interpret common pedestrian behaviour - e.g. stepping onto the road, making hand gestures or gazing at the vehicle - and react accordingly. Their interpretation and corresponding response need to be communicated in the visualisation.

\subsection{Limitations}

Among 18 participants, we had 14 international participants who recently moved to or had stayed for less than 5 years in Sydney, where the study was conducted. Their crossing behaviour may have been influenced by their cultural background and norms from their home cities. This represents a factor that we did not control in our study. 


VR has been shown to be a useful testing medium for interaction between pedestrian participants and AVs. 
However, using a simulated environment meant that we did not consider real-world environmental factors such as sounds, changing lighting conditions, weather conditions, other pedestrians, and vehicles - all conditions that may influence pedestrians' behaviours \cite{Pillai2017}. Further, there was a spatial difference between VR and the real world: 5 participants (P2, P5, P6, P9, P11) displayed a loss of balance when stepping down to the road for the first time in the VR environment as they expected a step when they were actually moving on a flat surface. This spatial difference may have affected our participants' behaviour to some extent. 
Similarly, the requirement of wearing a VR headset may have had an effect on how people moved and behaved in the VR environment. In our study, only one participant expressed that the VR headset felt uncomfortable. We were able to address this potential issue to some extent by getting participants to take the headset off between scenarios and sitting down in a comfortable booth for the post-scenario interviews. 


\section{Conclusions and Future Work}
In this research, we designed a novel projection-based design solution to enable AV-to-pedestrian communication and evaluated it with participants in VR with four street-crossing scenarios, three of which involved varied degrees of sensor malfunction. The evaluation gave us insights into people's understanding of communication messages and their behaviour around autonomous vehicles. Based on the findings from the evaluation, we proposed seven design recommendations for projection-based AV-to-pedestrian communication systems (see Table \ref{tbl:design-recommendations}).

In summary, our research offers three contributions for future work on AV-to-pedestrian communication:

\begin{enumerate}
\item We explored a novel projection-based design solution for communication between AVs and pedestrians.
\item We contributed insights on how people understand and interpret projection-based visual communication signals, including malfunction (non-detection) scenarios.
\item We presented a series of design recommendations to guide future design exploration.
\end{enumerate}

Future studies can build on our work to investigate scenarios that involve multiple vehicles and pedestrians at complex crossings. Such studies might help to refine and expand our initial set of design recommendations for this emerging field of research. In particular, our findings highlight three areas for future exploration. First, in the context of projections, solutions for diverse road surfaces, weather conditions and lighting conditions could be considered. Second, a targeted investigation of malfunctions (leading to non-detection) and their long-term influence on (over-)trust and behaviour of pedestrians could reveal additional design considerations and valuable insights for the design of future AVs. Third, as visualising a car's intent can add important information for pedestrians, future research could investigate how long in advance intentions should be visualised by an AV.

\newpage





%
\bibliographystyle{ACM-Reference-Format}
\bibliography{sigchi-latex-proceedings/references,sigchi-latex-proceedings/references-manual}

\end{document}